\documentclass[twocolumn,showpacs,aps,prd,amsmath,amssymb,superscriptaddress,nofootinbib,nobibnotes,floatfix]{revtex4-1}

\usepackage{graphicx}
\usepackage{float}
\usepackage{bm}
\usepackage[hidelinks]{hyperref}
\usepackage{color}

\usepackage[utf8]{inputenc}
\usepackage[T1]{fontenc}
\usepackage{slashed}

\usepackage{enumerate}

\usepackage{mathalfa,amssymb,amsthm}
\usepackage{amsmath}

\newcommand{\ssec}[1]{\section{#1}}

\begin{document}

\title{Dynamics of spontaneous black hole scalarization
   and mergers in Einstein-scalar-Gauss-Bonnet gravity
   }
\author{William E. East}
\email{weast@perimeterinstitute.ca}
\affiliation{%
Perimeter Institute for Theoretical Physics, Waterloo, Ontario N2L 2Y5, Canada.
}%
\author{Justin L. Ripley}
\email{lloydripley@gmail.com}
\affiliation{%
DAMTP,
Centre for Mathematical Sciences,
University of Cambridge,
Wilberforce Road, Cambridge CB3 0WA, UK.
}%
\begin{abstract}
We study the dynamics of black holes in Einstein-scalar-Gauss-Bonnet
theories that exhibit spontaneous black hole scalarization using recently
introduced methods for solving the full, non-perturbative equations of
motion.  For one sign of the coupling parameter, non-spinning vacuum black
holes are unstable to developing scalar hair, while for the other,
instability only sets in for black holes with sufficiently large spin.
We study scalarization in both cases,
demonstrating that there is a range of parameter space where the theory
maintains hyperbolic evolution and for which the instability saturates in a
scalarized black hole that is stable without symmetry assumptions.
However, this parameter space range is
significantly smaller than the range for which stationary
scalarized black hole solutions exist. We show how different choices for the
subleading behavior of the Gauss-Bonnet coupling affect the dynamics of the
instability and the final state, or lack thereof.  Finally, we present mergers of binary
black holes and demonstrate the imprint of the scalar hair in the
gravitational radiation.
\end{abstract}
\maketitle
\ssec{Introduction}%
In recent years, the observations of black holes (BHs) through
electromagnetic and gravitational waves
have furnished new opportunities to
test our understanding of gravity
(see, e.g. \cite{TheLIGOScientific:2016src,
   Yunes:2016jcc,
   Baker:2017hug,
   Abbott:2018lct, 
   Isi:2019aib, 
   Abbott:2020jks, 
   Isi:2020tac, 
   Psaltis:2020lvx, Volkel:2020xlc, Kocherlakota:2021dcv,
   Okounkova:2021xjv}).
However, in order to perform model selection
tests of General Relativity (GR) with these observations, one needs accurate
predictions for modified gravity theories in the strong field and dynamical
regime, which in these cases, is an outstanding theoretical problem.

A interesting class of theories to test against GR is Einstein-scalar-Gauss-Bonnet 
(ESGB) gravity.
Variants of ESBG give rise to BH solutions with scalar hair, 
hence can differ qualitatively from GR in the strong field regime (e.g. in
BH mergers) while still passing weak field tests.
Here we focus on versions of ESGB where GR solutions with vanishing scalar field
remain solutions of the modified theory, but which in some circumstances are
unstable to perturbations in the scalar field
(this contrasts with linearly-coupled ESGB, studied
extensively~\cite{Sotiriou:2014pfa,Benkel:2016rlz,Benkel:2016kcq,
Witek:2018dmd,Okounkova:2019zep,Okounkova:2020rqw,Ripley:2019irj,
Ripley:2019aqj,Delgado:2020rev,Sullivan:2020zpf},
where stationary BHs always form a scalar cloud.)
Recently, stationary scalarized BH solutions to the full equations
of motion were constructed in this class of theories~\cite{Doneva:2017bvd,Silva:2017uqg,
Minamitsuji:2018xde,Silva:2018qhn,
Dima:2020yac,Herdeiro:2020wei,Berti:2020kgk}.
These studies show that for
particular ranges of mass and spin, set by the magnitude and
sign of the Gauss-Bonnet (GB) coupling, the scalarized
BHs can differ significantly from their GR
counterparts (e.g. the BH spacetime can have $~20\%$ of its mass 
in the scalar cloud), likely impacting
gravitational wave observations.

These scalarized BH solutions 
are plausibly the end state of
the linear scalarization instability
of vacuum BHs of sufficiently small mass 
(in comparison to the GB coupling length scale)~\cite{Doneva:2017bvd,Silva:2017uqg} 
or sufficiently high spin 
(referred to as spin-induced scalarization)~\cite{Dima:2020yac},
which would provide a formation channel.
However, the scalarization process is poorly understood,
particularly in the spinning BH case,
where not even the stability of the scalarized solutions is known.
The dynamics in these theories have only been studied
by treating the scalar field as a test on a GR background, 
both for isolated~\cite{Dima:2020yac,Doneva:2021dqn,Kuan:2021lol} and binary BHs~\cite{Silva:2020omi}
(with the expection of a nonlinear study of Ref.~\cite{Ripley:2020vpk} which
considered a different ESGB variant than considered here in spherical symmetry).
A major challenge in studying the nonlinear dynamics 
for these theories has been in finding a
well-posed scheme for the ESGB equations of motion (EOM).
Here, we build on the methods of Ref.~\cite{East:2020hgw}, where we
demonstrated the feasibility of finding full binary BH solution solutions in
linearly-coupled ESGB gravity using the modified generalized harmonic
(MGH) formulation \cite{Kovacs:2020pns,Kovacs:2020ywu}.
Using these methods, we study the nonlinear development and saturation
of the scalarization instability for two variants of ESGB gravity,
working with initially nearly vacuum (binary) BH solutions with
a small scalar field perturbation. We find a range
of parameters where this initial data
leads to the formation of a stationary scalarized
BH. However, we also find that in a significant portion of the parameter space,
the predictability of the theory (i.e. the hyperbolicity of the equations
of motion) breaks down during
scalarization, even when stationary scalarized BH solutions exist.  
By studying binary BH mergers,
we show that even when restricted to this parameter space,
BH scalarization can have a significant impact on the
resulting gravitational waves. 

\ssec{Spontaneous BH scalarization in ESGB gravity
   \label{sec:esgb_spontaneous_scalarization}
}%
We first briefly review the theories we consider, 
and the heuristic arguments
behind why BHs may be dynamically unstable to scalar
field perturbations in these theories.

The action for ESGB gravity is
\begin{align}
\label{eq:the_action}
    S
    =
    \frac{1}{8\pi}
  \int d^4x\sqrt{-g}
\left(
        \frac{1}{2}R
    -\frac{1}{2}\left(\nabla\phi\right)^2 	
    +	\beta\left(\phi\right)\mathcal{G}
    \right)
    ,
\end{align}
where
$
\mathcal{G}
=
   R^2
-  4R_{\mu\nu}R^{\mu\nu}
+  R_{\alpha\mu\beta\nu}R^{\alpha\mu\beta\nu}
$ is the GB scalar, and we use geometric units with $G=c=1$ here 
and throughout.
ESGB gravity appears in the low-energy effective actions for certain
string theories \cite{Zwiebach:1985uq,Gross:1986mw} and more generally
captures the leading order scalar-tensor interactions 
expected in an effective gradient expansion of the Einstein
equations \cite{Weinberg:2008hq,Kovacs:2020pns}.
We will consider two different classes of scalar
GB coupling that allow for spontaneous BH scalarization:
\begin{subequations}
\begin{align}
\label{eqn:polynomial_coupling}
   \beta(\phi)
   =&
   \frac{\lambda}{2}\phi^2+\frac{\sigma}{4}\phi^4 
   ,\\
\label{eqn:exponential_coupling}
   \beta(\phi)
   =&
   \frac{\lambda_e}{6}\left (1-e^{-3\phi^2}\right) 
   ,
\end{align}
\end{subequations}
where $\lambda$, $\sigma$, and $\lambda_e$ are constants.
Note that these agree to leading order in $\phi^2$ when
$\lambda=\lambda_e$.
We choose these because scalarized BH solutions with these
couplings have been constructed and studied 
\cite{Doneva:2017bvd,Silva:2017uqg,
Minamitsuji:2018xde,Silva:2018qhn,
Dima:2020yac,Herdeiro:2020wei,Berti:2020kgk}.
The first captures the leading and first subleading term
for a coupling invariant under $\phi\rightarrow-\phi$,
while the second is one
particular higher-order completion with this property.

To review the idea behind spontaneous scalarization,
we only need to consider the scalar field equation of motion:
\begin{align}
\label{eq:scalar_field_eom}
   \Box\phi
   +
   \beta^{\prime}\left(\phi\right)\mathcal{G}
   =
   0
   , 
\end{align}
where, expanding around $\phi=0$, 
$\beta^{\prime}(\phi)=\beta^{\prime\prime}(0) \phi + \mathcal{O}(\phi^2)$.
Provided $\beta^{\prime\prime}\mathcal{G}>0$, this
term will act like a tachyonic mass, and for small enough BH 
masses, there will be a linear instability
which could potentially give rise to a stable scalarized
BH solution.
There are two possibilities for $\mathcal{G}$ and $\beta''$ to have
the same sign.
For non-spinning and slowly spinning BHs, $\mathcal{G}>0$
everywhere exterior to the BH, so in order to 
see spontaneous scalarization one needs $\beta''>0$
\cite{Doneva:2017bvd,Silva:2017uqg,
Minamitsuji:2018xde,Silva:2018qhn}.
For rapidly enough rotating BHs, $\mathcal{G}$ is no
longer positive definite, which allows for \emph{spin induced
spontaneous scalarization} if $\beta''<0$
\cite{Dima:2020yac,Herdeiro:2020wei,Berti:2020kgk}.

\ssec{Methodology\label{sec:methodology}}%
We numerically evolve the full ESBG EOM
using the MGH
formulation~\cite{Kovacs:2020pns,Kovacs:2020ywu}
as described in Ref.~\cite{East:2020hgw}.
We use similar choices for the gauge, numerical parameters, etc.\ as in
Ref.~\cite{East:2020hgw},
except that we find the scalarized BHs evolved here also benefit from 
the addition of long wavelength constraint damping obtained by setting
$\rho=-0.5$ in Eq.~(2) of Ref.~\cite{East:2020hgw}.

For initial data, we start from single or binary vacuum BH solutions
(the latter constructed as in Ref.~\cite{East:2012zn}) 
with a small Gaussian scalar perturbation centered on the BH(s).
For most cases presented here, we use an initial amplitude
of $\phi_0=0.01$,
though we have verified smaller
amplitudes give the same results, and 
that the error induced by not solving
the constraint equations including the perturbation
is negligible. See the Supplementary Material 
(which cites the references
\cite{Minamitsuji:2018xde,Silva:2018qhn,Doneva:2017bvd,Silva:2017uqg,
East:2020hgw,Ripley:2020vpk,Hayward:1994bu})
for details on resolution, convergence, and the
exact form of the initial perturbation.

We use many of the same diagnostics as in 
Ref.~\cite{East:2020hgw}, which we briefly review here.
We measure the gravitational radiation by extracting the Newman-Penrose
scalar $\Psi_4$, and use this to calculate an associated gravitational
wave luminosity $P_{\rm GW}$. We also measure the flux of energy in the scalar field $P_{\rm SF}$. 

During the evolution, we track any apparent horizons present at a given time,
and measure their areas and associated angular momentum $J_{\rm BH}$. From this
we compute a BH mass $M_{\rm BH}$ via the Christodoulou formula.  We will refer
to the mass that lies outside the BH horizon(s)---which, to a good
approximation, can be attributed to the scalar cloud---as
$M_{\phi}\equiv M-M_{\rm
BH}$, where $M$ is the global Arnowitt-Deser-Misner (ADM)
mass of the spacetime, and similarly define
an angular momentum $J_{\phi}\equiv J-J_{BH}$.
Finally, we compute the scalar charge from the
asymptotic behavior of the scalar field at large $r$: $\phi=Q_{\rm
SF}/r+\mathcal{O}(1/r^2)$, where we have fixed that $\phi \rightarrow 0$ at
spatial infinity with our initial conditions.

In addition to evolutions with the MGH 
formulation, we also present evolutions of BHs in spherical
symmetry, using the formalism and code described in
Ref.~\cite{Ripley:2020vpk}.  We do this in order to determine exactly
where the hyperbolicity of ESGB breaks down in spherical symmetry for various
choices of coupling, which we can compare to the MGH 
evolutions without symmetry assumptions.  Spherically symmetric spacetimes are
not only computationally less expensive (and thus we are able to
systematically scan the parameter space),
but there is less gauge
ambiguity in determining when the EOM are
hyperbolic (although see Ref.~\cite{Reall:2021voz} for a recently
introduced formalism for generic backgrounds). 

\ssec{Black hole scalarization and saturation}
We begin by considering the scalarization of isolated 
non-spinning and spinning BHs using our MGH code \cite{East:2020hgw}.
Our main result is that we
find a range of parameters for both signs of $\beta''(0)$ 
where the scalarization instability stability saturates in the formation
of a stable BH with scalar hair containing up to a few percent of the total mass.  

\begin{figure}
\begin{center}
    \includegraphics[width=\columnwidth,draft=false]{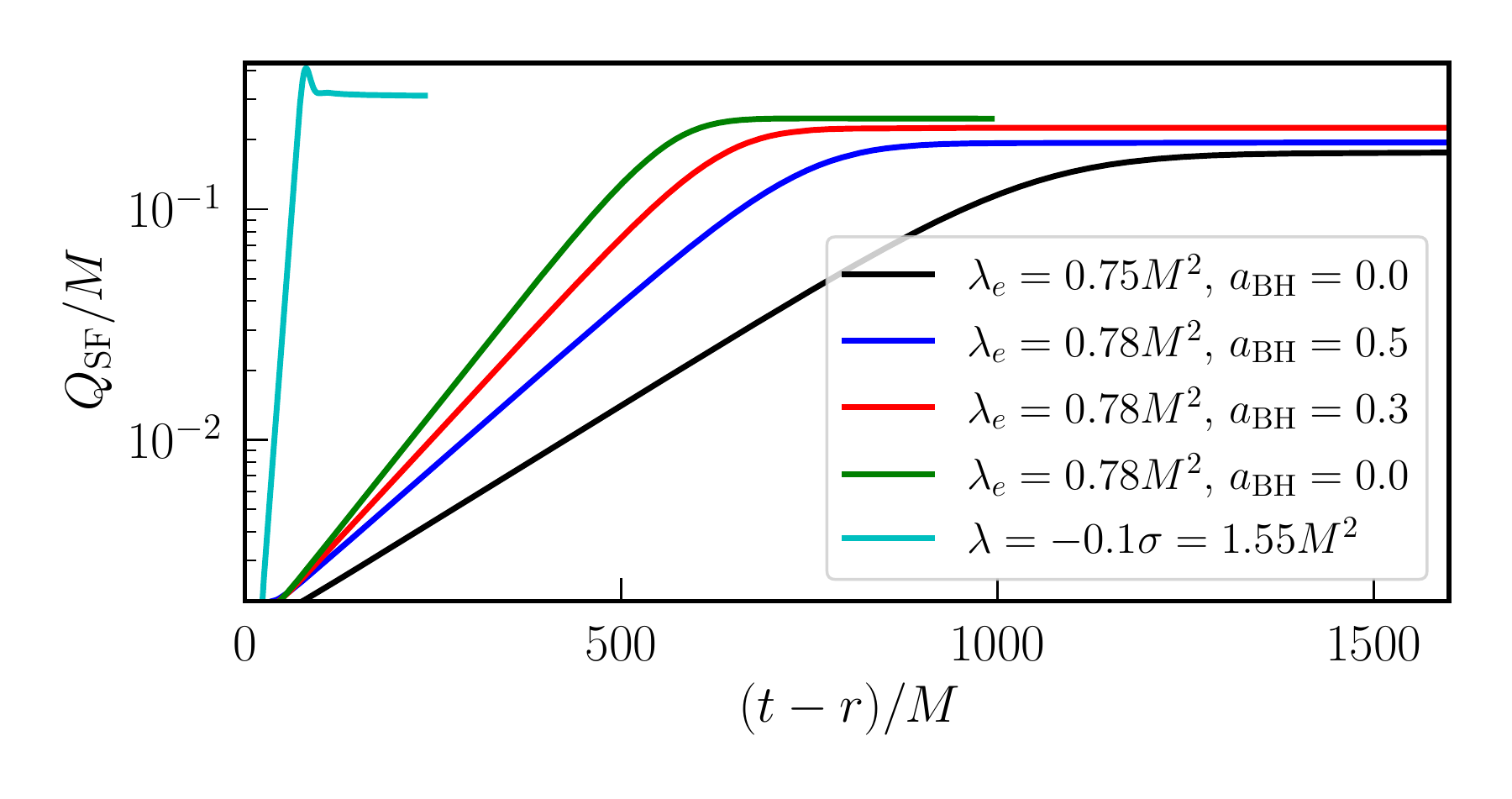}
    \includegraphics[width=\columnwidth,draft=false]{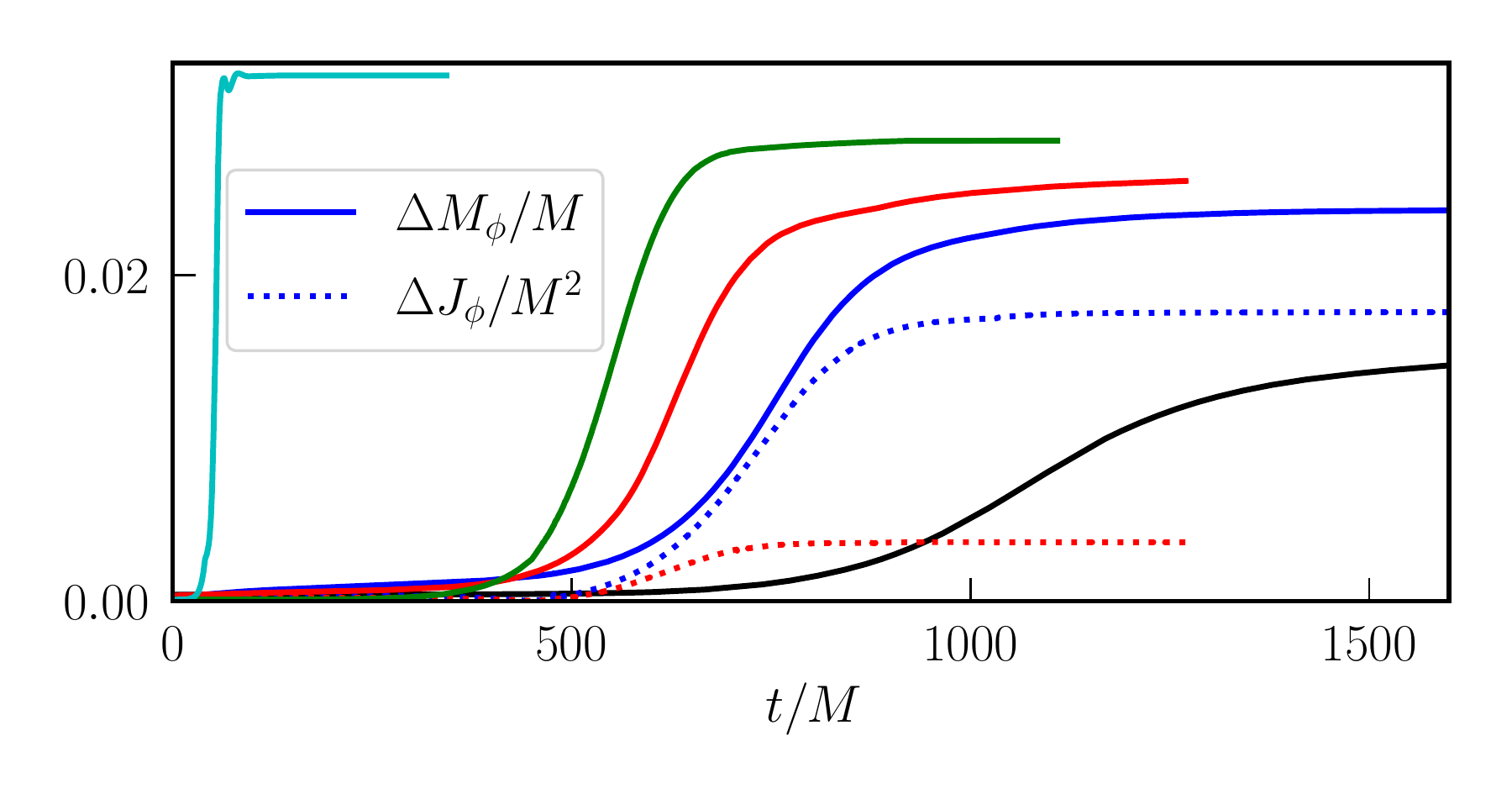}
\end{center}
\caption{
Scalarization of BHs with positive GB coupling.  Top: The scalar charge
measured at a large distance as a function of look-back time.  
Bottom: The amount
of mass (solid lines) and, for the cases with spin, angular momentum (dashed
lines) going from the BH into the scalar cloud and radiation.  The legend in
the top panel applies to the bottom as well.
\label{fig:pos_lam_scalarize}
}
\end{figure}

Considering first $\beta''(0)>0$, we show the dynamics of BH scalarization
for several cases in Fig.~\ref{fig:pos_lam_scalarize}. Following an
exponential growth phase, where the BH develops scalar charge at the
expense of losing mass, we find that the instability eventually saturates
and settles to a nearly stationary
scalarized BH solution.
(We note that the area of BHs can decrease in theories like ESGB
that violate
the Null Convergence Condition, see e.g. \cite{Ripley:2019irj}.)
Increasing $\lambda_e$ results in higher instability rates
and more massive scalar clouds at saturation.
At fixed coupling, considering non-zero BH spin
decreases the instability rate and cloud mass.  
In addition to results with the exponential
coupling [Eq.~(\ref{eqn:exponential_coupling})], in
Fig.~\ref{fig:pos_lam_scalarize} we also show one case with the coupling
given by Eq.~(\ref{eqn:polynomial_coupling}) and $\lambda=-\sigma/10=1.4M^2$.
(We note that with $\sigma=0$,
spherical scalarized BHs are radially
unstable~\cite{Blazquez-Salcedo:2018jnn}.)  
In this case, a similar amount of mass goes into the scalar
cloud compared to the exponential coupling with $\lambda_e=0.78M^2$, but
the instability happens much faster and initially overshoots, e.g.,
the final scalar charge (see top panel of Fig.~\ref{fig:pos_lam_scalarize}).

We were unable to obtain hyperbolic evolutions through
saturation for non-spinning BHs with positive $\lambda_e$ or
$\lambda=-\sigma/10$ much higher than the above mentioned values.
As we discuss in the next section,
this is because we are approaching the regime
where the asymptotic hyperbolicity of the theory breaks down.

\begin{figure}
\begin{center}
    \includegraphics[width=\columnwidth,draft=false]{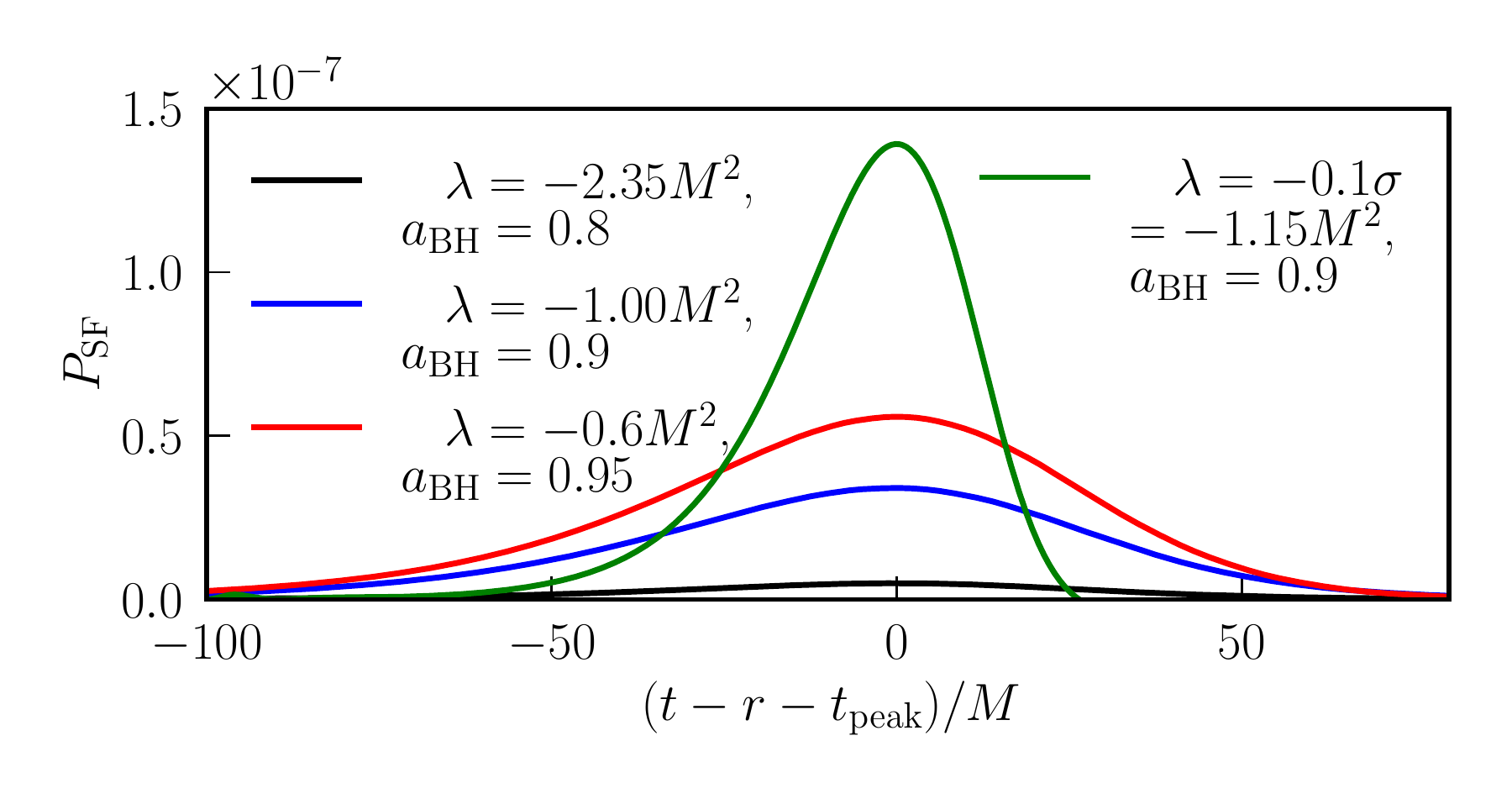}
    \includegraphics[width=\columnwidth,draft=false]{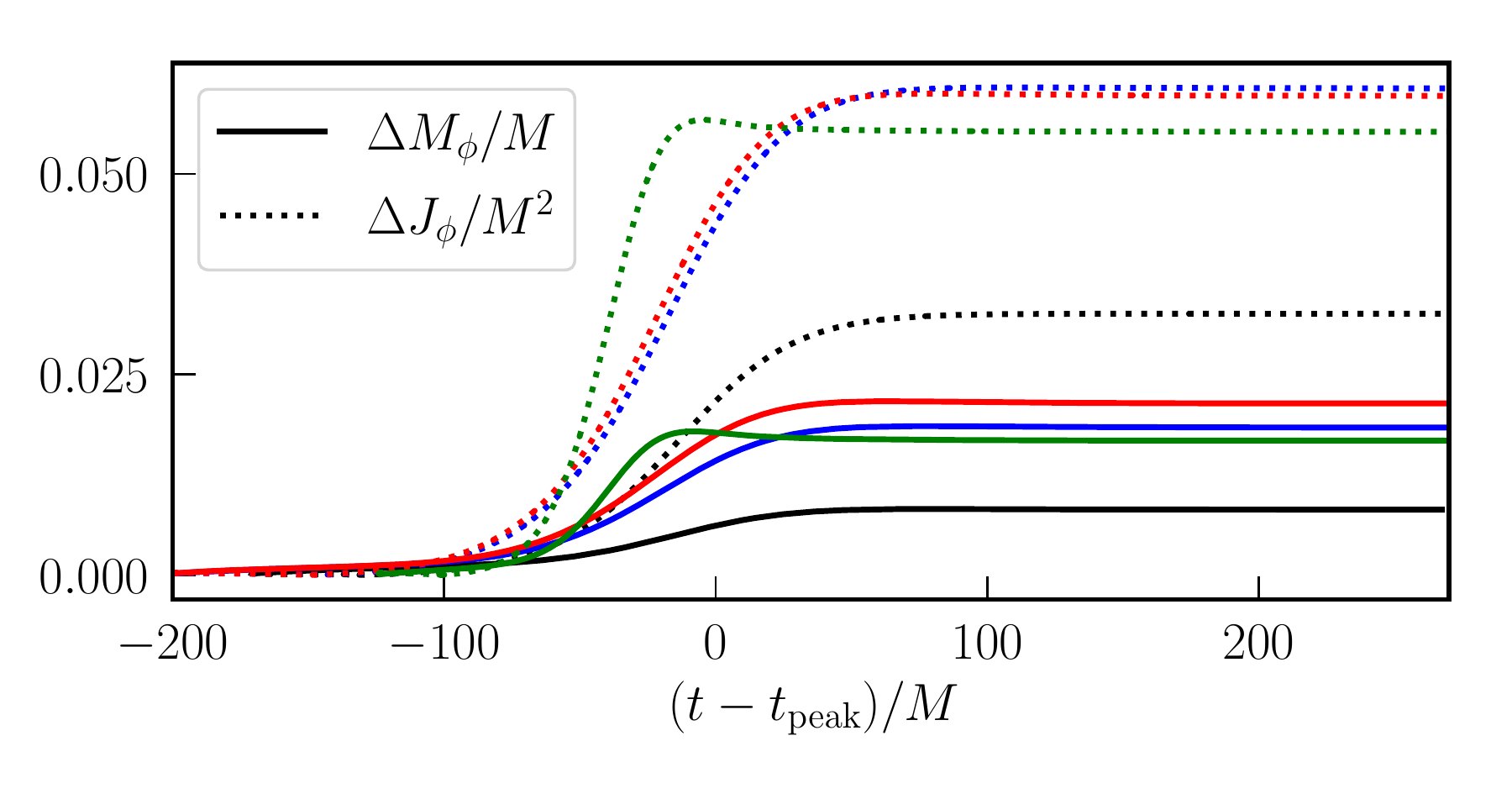}
\end{center}
\caption{
Scalarization of BHs with negative GB coupling.  Top: The scalar energy flux.
Bottom: The amount of mass (solid lines) and angular momentum (dashed lines)
going from the BH into the scalar cloud and radiation.  The time axis is
shifted according to (the look-back time of) the peak of the radiation.  The
legend in the top panel applies to the bottom as well.
\label{fig:neg_lam_scalarize}
}
\end{figure}

We also consider spin-induced BH scalarization, which occurs when
$\beta''(0)<0$, finding similar results. In
Fig.~\ref{fig:neg_lam_scalarize}, we show several different cases where
the initial BH spin ranges from $a_{\rm BH}=0.8$ to $a_{\rm BH}=0.95$ with
different values of the coupling given by Eq.~(\ref{eqn:polynomial_coupling}) 
(with $\sigma=0$ unless otherwise noted).  We again find a range of parameters
where the scalarization instability saturates and leads to the formation
of a stationary BH solution with up to a few percent of the mass of the BH
converted to scalar hair. For higher values of spin, the instability sets
in at lower values of the coupling. 
We find the gravitational waves from the scalarization process
to be negligible, but the scalar radiation increases with the instability rate 
(see top panel of Fig.~\ref{fig:neg_lam_scalarize}; the time dependence
of $Q_{\rm SF}$ from these cases is similar to the $\beta''(0)>0$ cases).
Though not shown, we also considered
an exponential coupling up to $\lambda_e=1.05M^2$ for $a_{\rm BH}=0.9$ and
found similar results. 
Again, we were not able to obtain numerical evolutions for significantly
higher values of the coupling than shown in Fig.~\ref{fig:neg_lam_scalarize}
for the given BH spin values. That is, the breakdown in the evolution
occurs for lower coupling values for higher BH spins. 

\ssec{Limits on scalarized BHs from hyperbolicity}
\label{sec:results_spherical_sym}
For the case of spherically symmetric BHs (i.e. using the code in
Ref.~\cite{Ripley:2020vpk}), we will now explicitly demonstrate that the reason
we are unable to follow the dynamical formation of BHs with scalar hair
containing more than a few percent of the total mass 
is because the hyperbolicity of the theory
breaks down (and we conjecture that something similar happens beyond spherical
symmetry). 

\begin{figure}
    \centering
    \includegraphics[width=0.8\columnwidth,draft=false]{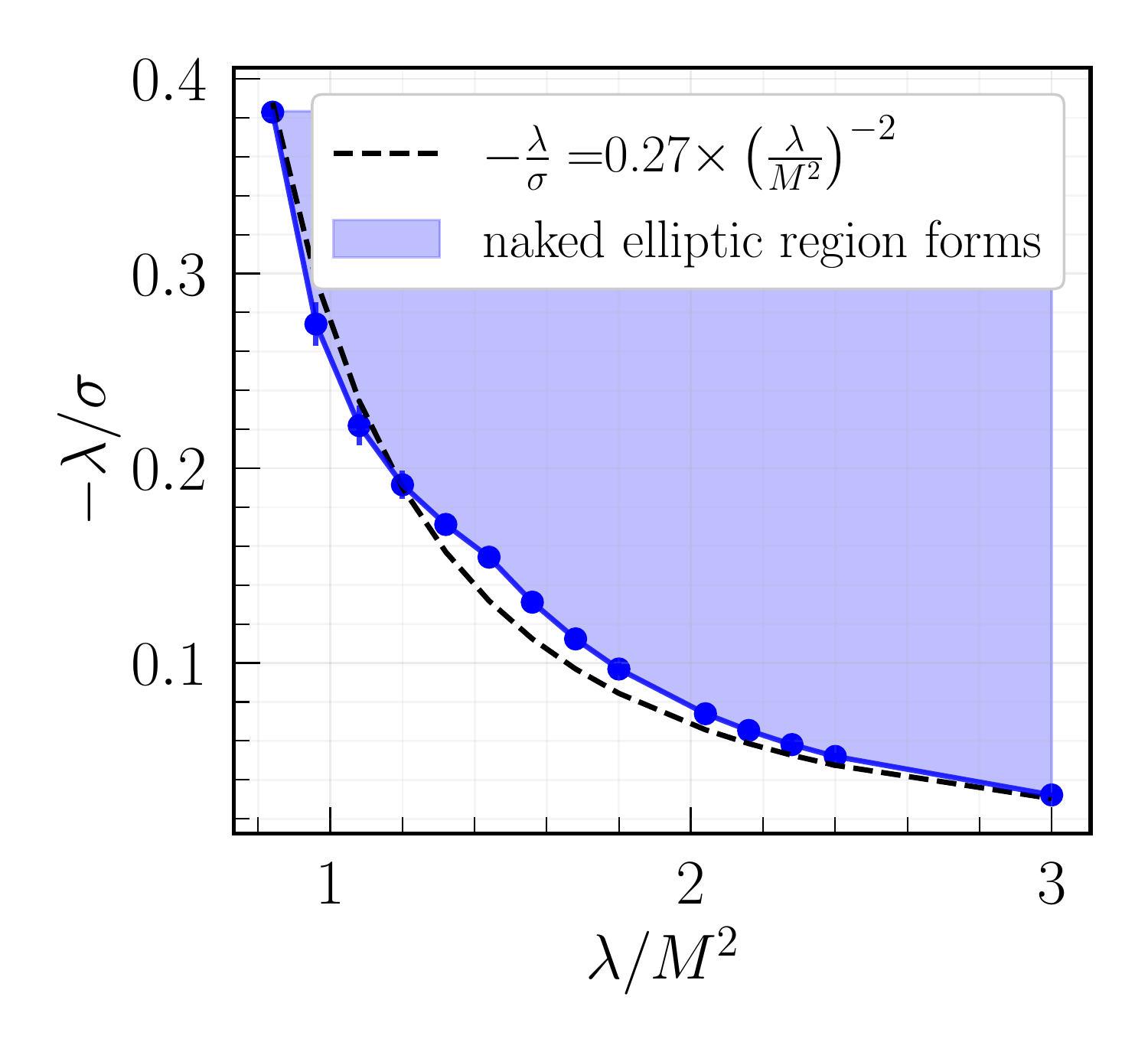}
    \centering
    \includegraphics[width=0.8\columnwidth,draft=false]{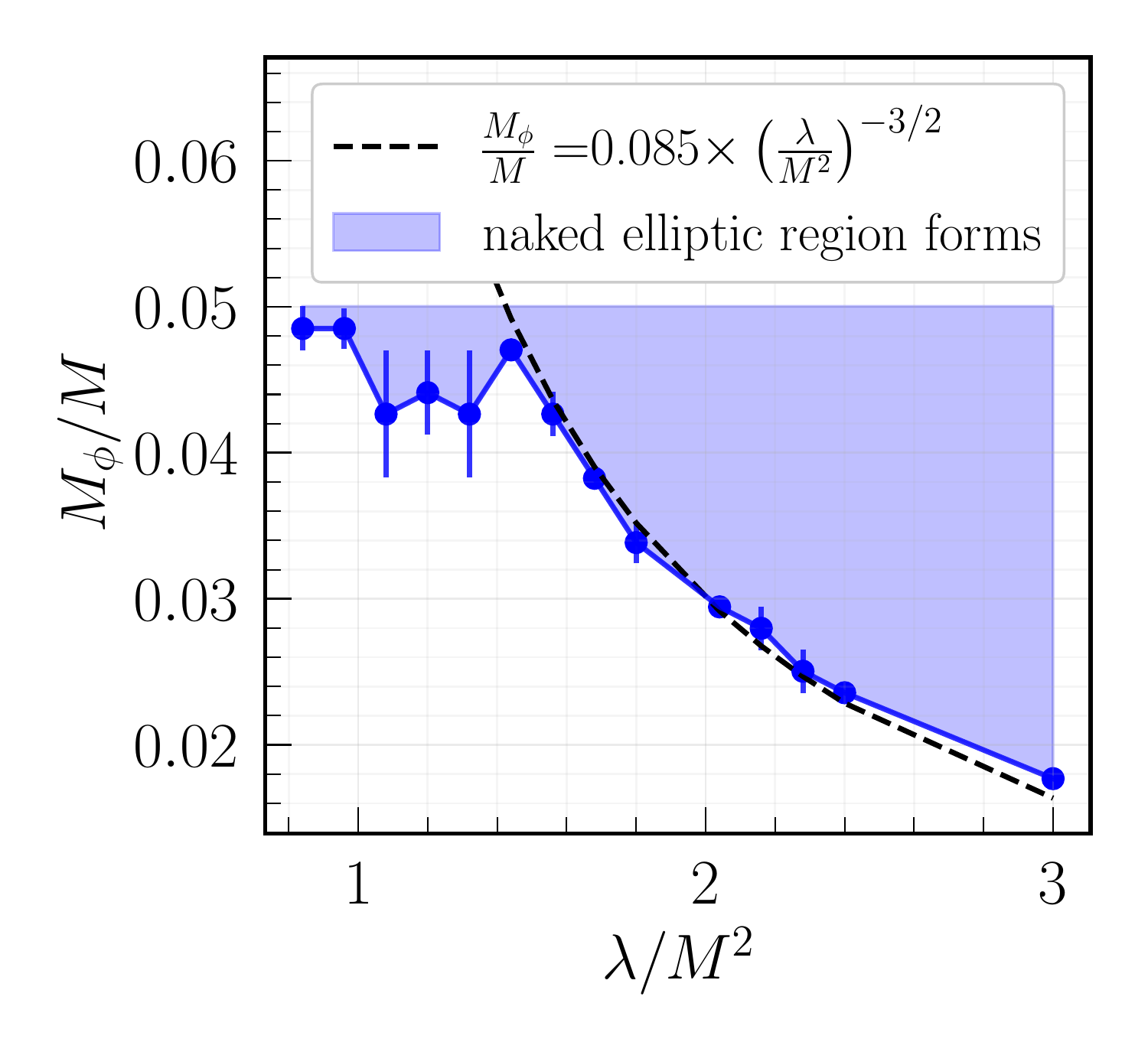}
\caption{
   The region of parameter space (shaded blue) 
   where an elliptic region forms outside the BH
   during scalarization with spherical symmetry enforced,
   for the polynomial coupling Eq.~\eqref{eqn:polynomial_coupling}.
   We show this region in terms of the ratio
   of the quadratic-to-quartic coupling $\lambda/\sigma$ (top),
   and the maximum fraction of the global mass liberated from the
   BH $M_{\phi}$ at the extremal value of $\sigma$ (bottom),
   for a given value of $\lambda/M^2$.
   The black dashed lines give least-squared monomial fits
   to these quantities.
   The error bars are the difference in the extremal $\sigma$
   and $M_{\phi}$ values computed at two different
   resolutions.
\label{fig:ellp_hyperbolic_line}
}
\end{figure}

For the polynomical coupling,
Eq.~\eqref{eqn:polynomial_coupling},
we fix $\lambda/M^2$, and then find the value of $\sigma$
for which the theory becomes elliptic.
As the former is made larger, the latter must be
made more and more negative in order
to control the magnitude of the coupling
$\beta'$ at saturation.
We plot the dividing line in parameter space in
Fig.~\ref{fig:ellp_hyperbolic_line}.
We find that for a given $\lambda\gtrsim0.84$
the minimum absolute value of $\sigma$ that is necessary for
hyperbolic evolution through scalarization is
\begin{equation}
\label{eq:hyperbolicity_relations}
   \frac{\sigma}{M^2}
   \lesssim
   -3.7 \times \left(\frac{\lambda}{M^2}\right)^3
   .
\end{equation}

The bottom panel of Fig.~\ref{fig:ellp_hyperbolic_line}
shows that the maximum amount of energy liberated from the BH 
and put into the scalar cloud by the instability
is always $\lesssim 5\%$, and the maximum falls of like 
$M_{\phi}\lesssim8.5\times 10^{-2}\times \lambda^{-3/2} M^4$
for $\lambda/M^2>1.5$. This may be connected to the fact that,
as illustrated above, larger values of $\lambda$ tend to 
initially overshoot the scalarized solution (as opposed to smaller
values of $\lambda$ which smoothly saturate) and hence may
more easily violate hyperbolicity dynamically.

For the case of positive exponential coupling,
Eq.~\eqref{eqn:exponential_coupling}, we find the dividing coupling between
hyperbolic and elliptic evolution for spherical scalarization 
is $\lambda_e=0.834M^2$.  This is within $\sim10\%$ of
the value one would obtain by neglecting $\mathcal{O}\left(\phi^6\right)$ terms
in this coupling and using Eq.~\eqref{eq:hyperbolicity_relations} with
$\sigma=-3\lambda$. By contrast, the analysis of
Ref.~\cite{Blazquez-Salcedo:2018jnn} concluded that scalarized BH solutions
with up to $\lambda_e\approx 8.55 M^2$ were radially stable (and similar
results were found for the quadratic-quartic
coupling~\cite{Minamitsuji:2018xde,Silva:2018qhn}.) 

Comparing these results to
the MGH evolutions that do not explicitly enforce spherical symmetry,
we see that the values of $\lambda_e$ used for the former
are within $10\%$ of the maximum value that retains hyperbolicity in spherical
symmetry (and within $\sim15\%$ for $\lambda=-0.1\sigma$). This difference is
likely just due to the more limited numerical resolution used for the MGH
evolutions, as approaching extremality, the narrowing region between the elliptic region
and the horizon becomes more and more difficult to resolve. 

Going beyond spherical symmetry to the case of spin-induced scalarization, we
do not have any definitive results on the breakdown of hyperbolicity
(only positive results establishing hyperbolicity for a range of parameters). However,
given the above, we can conjecture that the reason we were not able to evolve
significantly larger couplings is that, for this case as well, elliptic regions 
develop outside the BH horizon during scalarization, even in the regime
where stationary scalarized BH solutions exist.

\ssec{Head on collisions of scalarized BHs}
Given that we find that scalarized BHs with up to a
few percent of the total mass in the scalar cloud
can form in the regime where the theory is hyperbolic,
it is interesting to consider how this will affect a binary BH merger.
Here we focus on the case of the (axisymmetric)
head-on collision of an equal-mass binary, 
considering both a non-spinning binary (with $\lambda>0$)
and a spinning binary (with $\lambda<0$).

\begin{figure}
\begin{center}
    \includegraphics[width=\columnwidth,draft=false]{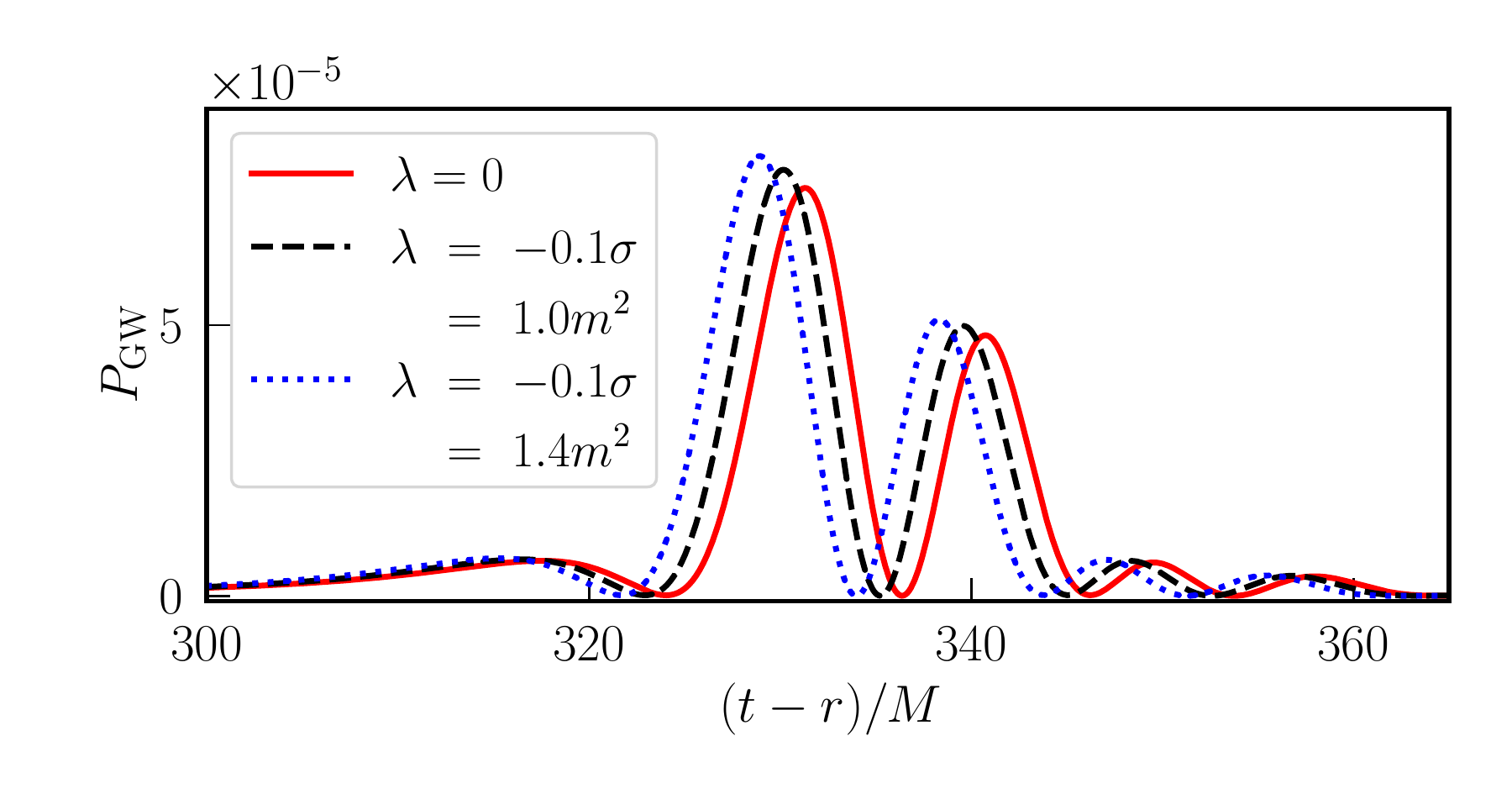}
    \includegraphics[width=\columnwidth,draft=false]{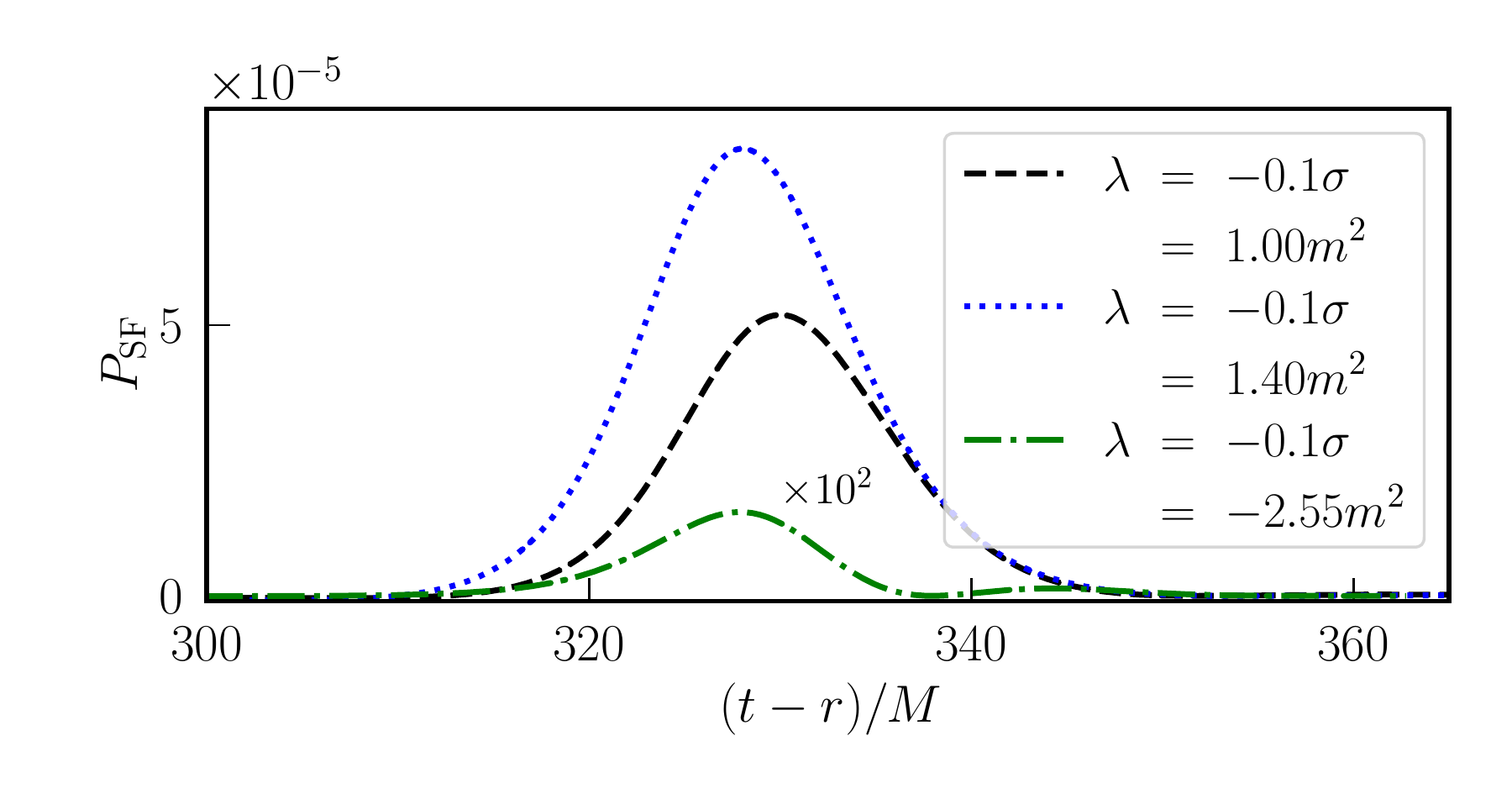}
\end{center}
\caption{
   The gravitational (top) and scalar (bottom) luminosity for the head-on
merger of equal mass BHs. For the $\lambda \geq 0$ cases (red, black, and
 blue curves), the BHs are non-spinning, while for the $\lambda<0$ case (green curve;
multiplied by $\times10^2$ to be visible), 
each member of the binary has dimensionless spin $a_{\rm BH}=0.8$ initially.
We see that the gravitational wave luminosity is
larger when two scalarized BHs collide, as compared to two vacuum
BHs in the theory.
}
   \label{fig:pos_lam_bhbh}
\end{figure}

In Fig.~\ref{fig:pos_lam_bhbh}, we plot the total gravitational wave
and scalar field luminosity for head-on black
hole collisions.  As described above,
we begin with two nearly-vacuum BH solutions, with a small scalar
field perturbation outside their horizons. With the couplings we consider,
the BHs subsequently spontaneously scalarize, and reach saturation
well before they collide.  For $\lambda=-0.1\sigma \gtrsim m^2$ (where $m$
is the mass of one of the binary constituents) we find that the
gravitational waves from the scalarized BH collisions have noticeably
larger amplitude compared to the GR case, and that the scalar luminosity
is comparable to the gravitational wave luminosity.
Due to the increased radiation with larger $\lambda$, the merger also happens slightly faster
for the same initial separation/velocity.

We also show a case with $\lambda=-0.1\sigma=-2.55m^2$ where (unlike the
above mentioned case) the binary constituents are both spinning with $a_{\rm
BH}=0.8$. Here, despite the fact that $\sim0.9\%$ of the mass is a
spin-induced scalar cloud prior to merger (compared to $\sim1.7\%$ and $\sim2.8\%$
for the above cases with $\lambda/m^2=1$ and 1.4, respectively), the scalar radiation from
merger is several orders of magnitude smaller than the positive $\lambda$
cases, and the gravitational wave luminosity (not shown) does not noticeably differ
from the GR case. Hence, it appears that larger BH spins are necessary to have
a strong impact on the binary BH collision for spin-induced scalarization
(though for an inspiral, the impact is likely greater).

\ssec{Discussion and Conclusion}%
In this work we have studied full, non-perturbative solutions to several ESGB
theories that exhibit spontaneous BH scalarization.  We studied theories where
the GB coupling was either a single-parameter exponential function, or a two
parameter quadratic function of $\phi^2$, and for both choices of sign, leading
to either mass or spin-induced scalarization.  Though these theories have
a multidimensional parameter space, which we have not fully explored,
we can infer
several general results.  We have shown that the
end state of the nonlinear evolution of the linear scalarization instability of
vacuum non-spinning and spinning
BHs~\cite{Doneva:2017bvd,Silva:2017uqg,Minamitsuji:2018xde,Silva:2018qhn}
results in the formation of a stable, scalarized BH, for a range of coupling
parameters and BH masses and spins.  However, we find that for a given set of
couplings, at large enough curvature scales, the theory can lose hyperbolicity
and, in contrast to the linearly coupled
theory~\cite{Ripley:2019hxt,Ripley:2019irj,East:2020hgw}, this breakdown occurs
at much lower values compared to the maximum values where stationary solutions
can constructed.  In particular, while stationary solutions can be constructed
with over $20\%$ of the total mass is attributable to the scalar cloud, here we
did not find any cases where this was greater than 5\%.  

Thus, in the quadratic (plus higher order) coupled ESGB theories considered
here, as for the linearly coupled case, there is a minimum mass for a stable BH
to form. In the former case, vacuum BHs below this limit are unstable, while in
the latter case they are explicitly non-stationary. But in either case, their
subsequent evolution will generically result in a breakdown of the
predictability of the theory.  Nevertheless, the valid range is still
interesting from the point of view of potentially impacting the
gravitational wave signal of a binary BH merger, as we have demonstrated for some
example head-on binary collisions.  

We note that, while here we studied the scalarization instability starting from
stationary BHs, instead of following the dynamical formation of an unstable BH
from collapsing matter (see Ref.~\cite{Kuan:2021lol} for such a calculation in
the spherically symmetric, test-field limit), we expect our results to apply to
such cases as well.  This is because BHs generically form with some matter
distribution falling within its Schwarzschild radius at relativistic speeds,
while the scalarization instability timescale is generally much longer than the
BH light-crossing time.

Finally, for larger GB couplings, the scalarization instability growth rate is
faster and there is a tendency to initially overshoot the final stationary
solution, resulting in more scalar radiation. However, maintaining
hyperbolicity during the evolution also requires larger higher order
corrections that reduce the coupling at large field values,   and the energy
scale at which this breakdown occurs becomes smaller for larger couplings. 

This work also shows another example where
the methods for numerically evolving Horndeski modified gravity theories
of Ref.~\cite{East:2020hgw},
based on the MGH formulation~\cite{Kovacs:2020pns,Kovacs:2020ywu},
work at larger couplings, where the deviations
from GR are significant. For future work, it would interesting to 
study the predicted gravitational wave signal from scalarized binary
BH mergers in these theories. 

\ssec{Acknowledgments}
We thank Thomas Sotiriou for discussions on ESGB  gravity and
spontaneous scalarization, and Max Corman for comments on our article.
W.E. acknowledges support from an NSERC Discovery grant.
This research was
supported in part by Perimeter Institute for Theoretical Physics. Research at
Perimeter Institute is supported by the Government of Canada through the
Department of Innovation, Science and Economic Development Canada and by the
Province of Ontario through the Ministry of Research, Innovation and Science.
This research was enabled in part by support provided by SciNet
(www.scinethpc.ca/) and Compute Canada (www.computecanada.ca).
Some of the simulations presented in this article were performed on
computational resources managed and supported by Princeton Research Computing,
a consortium of groups including the Princeton Institute for
Computational Science and Engineering (PICSciE)
and the Office of Information Technology's High Performance
Computing Center and Visualization Laboratory at Princeton University.
\bibliographystyle{apsrev4-1.bst}
\bibliography{mod_grav}

\appendix
\section{Translation to other conventions}
\label{app:conv}
In this article we considered two different classes of scalar
GB coupling:
\begin{subequations}
\begin{align}
\label{eqn:polynomial_coupling}
   \beta(\phi)
   =&
   \frac{\lambda}{2}\phi^2+\frac{\sigma}{4}\phi^4 
   ,\\
\label{eqn:exponential_coupling}
   \beta(\phi)
   =&
   \frac{\lambda_e}{6}\left (1-e^{-3\phi^2}\right) 
   ,
\end{align}
\end{subequations}
Spontaneous scalarization via the coupling
Eq.~\eqref{eqn:polynomial_coupling} was first studied
by Refs.~\cite{Minamitsuji:2018xde,Silva:2018qhn}.
To translate our results into their notation:
in Ref.~\cite{Minamitsuji:2018xde} 
$\lambda\leftrightarrow \eta/4$ and 
$\sigma\leftrightarrow \eta\alpha$,
and in Ref.~\cite{Silva:2018qhn} 
$\lambda\leftrightarrow \bar{\eta}/4$ and 
$\sigma\leftrightarrow\bar{\zeta}/2$.
The authors in Refs.~\cite{Doneva:2017bvd,Silva:2017uqg}
studied spontaneous scalarization for the coupling
Eq.~\eqref{eqn:exponential_coupling}.
Our $\lambda_e$ corresponds to their $\lambda^2/4$.
All of these studies considered
small perturbations about static Schwarzschild and scalarized
BH solutions to argue for the instability of the Schwarzschild
BH and stability of the scalarized BH solution.
\section{Numerical convergence and comparison 
of axisymmetry and 3D}
\label{app:num}
Here we present convergence test results for
the modified generalized harmonic code \cite{East:2020hgw}
and spherically symmetric code \cite{Ripley:2020vpk} that we use,
as well as compare results with different symmetries imposed.

\begin{figure}
\begin{center}
    \includegraphics[width=\columnwidth,draft=false]{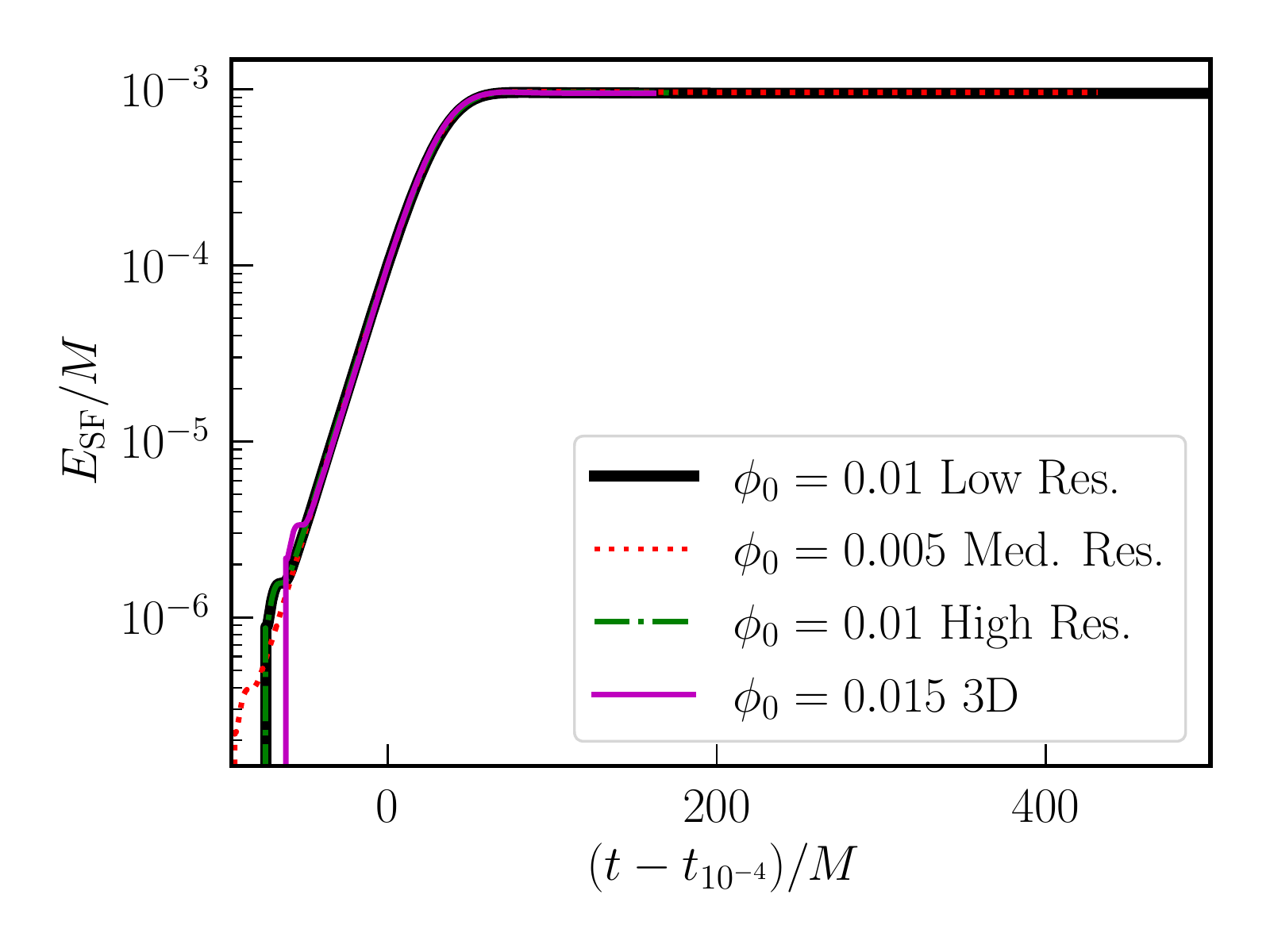}
    \includegraphics[width=\columnwidth,draft=false]{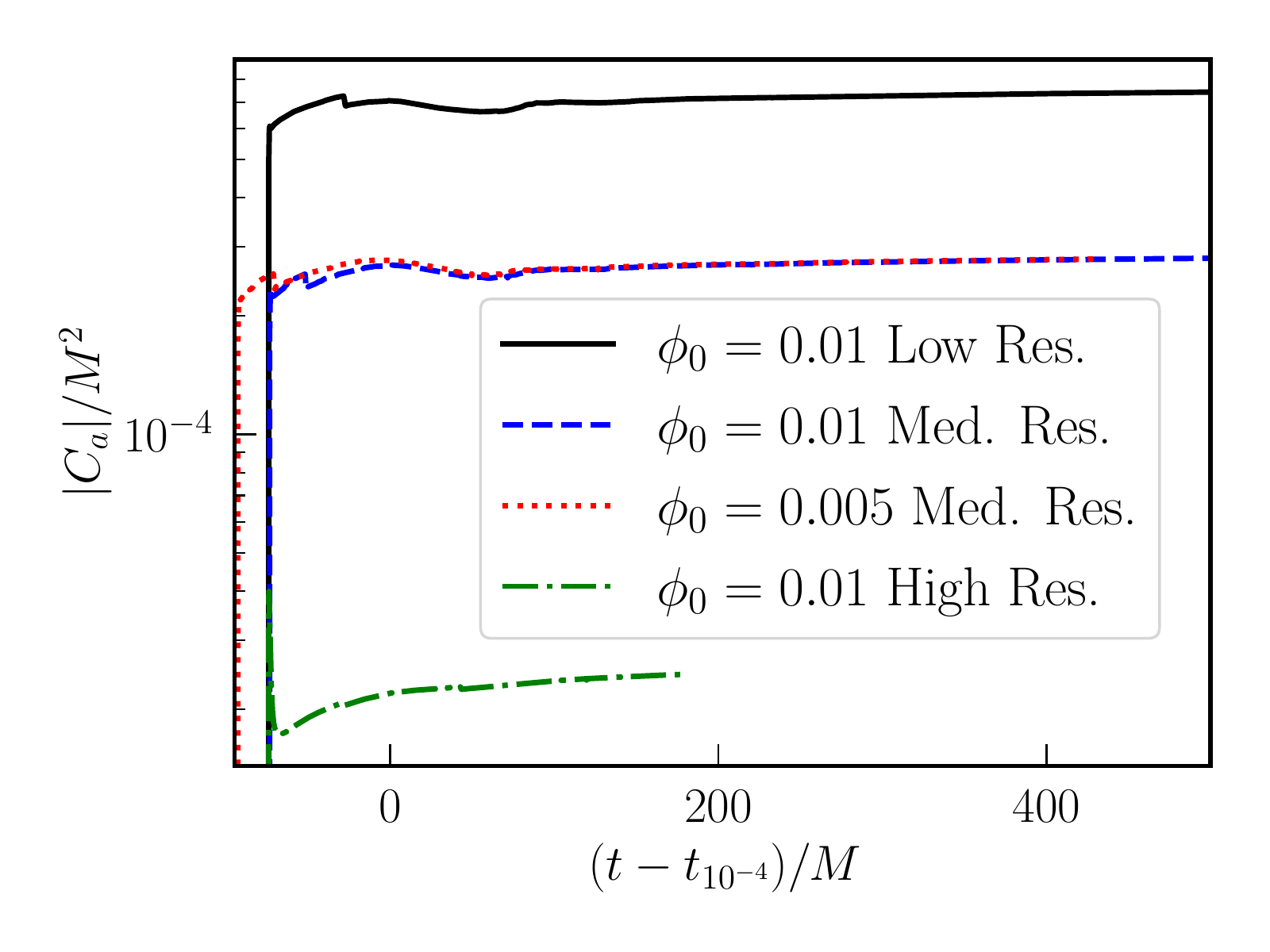}
\end{center}
\caption{
The scalar field energy (as measured from the canonical stress-energy tensor;
top panel) and violation of the MGH constraints
(bottom panel) for the scalarization of a BH with dimensionless spin $a_{\rm BH}=0.9$ and coupling
$\lambda=-\sigma/10=1.1M^2$. The initial scalar perturbation has amplitude
either $\phi_0=0.01$ or 0.005 for the axisymmetric cases, or
$\phi_0=0.015$ (and is equatorially offset from the BH) for the fully
3D case. 
A constant offset has been applied to the
time shown on the horizontal axis, so that the curves align
   when $E_{\rm SF}/M=10^{-4}$.
After a brief transient phase, we find the linear growth and nonlinear saturation to be the same.
We show results from three different numerical
resolutions, where the medium and high resolutions correspond to, respectively,
$\times 4/3$ and $\times 8/3$ as many points in each linear dimension compared to the
low resolution.  The decrease in the constraints with resolution is consistent
with approximately fourth order convergence.
\label{fig:scalarize_comp}
}
\end{figure}

In Fig.~\ref{fig:scalarize_comp}, we compare the spontaneous
scalarization of a spinning BH with $a_{\rm BH}=0.9$
and GB coupling $\lambda=-\sigma/10=1.1M^2$ for different
numerical resolutions and initial scalar perturbations. 
The medium resolution shown here uses 7 levels of mesh refinement with a refinement
ratio of $2:1$, and has a grid spacing of $dx\approx0.01M$ on the finest level,
which covers the BH. This is the default resolution for the results presented in this
work. For the initial perturbation, we use a Gaussian profile for the scalar
that is centered on the BH:
\begin{equation} 
\phi(t=0)=\phi_0\exp[-r^2_{\rm BH}/(2M_{\rm BH}^2)]\ .
\end{equation}
For most cases, we use an initial amplitude of $\phi_0=0.01$.  However, in
Fig.~\ref{fig:scalarize_comp} we also show a case with $\phi_0=0.005$ to
demonstrate that this is sufficiently far in the linear regime of the
instability, and that the constraint violation due to not including the
backreaction of the scalar in the initial data is smaller than the truncation
error of the highest resolution studies we considered.  We find that
the constraints converge to zero at the expected fourth order.

Finally, in the top panel of Fig.~\ref{fig:scalarize_comp}, we also show a case where we do
not enforce axisymmetry (in contrast to the other cases), and where
we choose the center of the initial Gaussian perturbation to be offset from the
BH by $M/4$ in the equatorial plane, in order to explicitly break axisymmetry.
We find that scalarization and saturation proceeds as in the axisymmetric case.

For the spherical symmetric evolutions, we do exactly solve the constraint equations for ESGB
gravity when constructing perturbed initial data.
As in Ref.~\cite{Ripley:2020vpk}
we use initial data that consisted of a small amplitude
scalar field bump profile outside of a Schwarzschild BH:
\begin{align}
   &\phi(t,r)\big|_{t=0}
   \nonumber\\
   =
   &\begin{cases}
      a_0\left(r-r_l\right)^2\left(r_u-r\right)^2
      \\
      \times\mathrm{exp}\left(
         -\frac{1}{(r_u-r)(r-r_l)}
      \right)
      &
      r_l<r<r_u
      \\
      0
      &
      \mathrm{otherwise}
   \end{cases}
   ,
\end{align}
where $r_u>r_l>2M$, where $M$ is the initial Misner-Sharp mass of the BH (see
\cite{Ripley:2020vpk} for more discussion).  We consider $r_l=2.4M$,
$r_u=5.2M$, and $a_0=10^{-2}$, so that the initial scalar field seed only adds
$\delta M_{\phi}/M\lesssim10^{3}$ to the total mass of the spacetime.  In
spherical symmetry, we compute the global mass by extracting the Misner-Sharp
mass at spatial infinity~\cite{Ripley:2020vpk}
(we note that in spherical symmetry the Misner-Sharp mass
is equal to the ADM mass at spatial infinity \cite{Hayward:1994bu}).
Note that we need $a_0\neq0$,
as the Schwarzschild solution is a stationary solution to the theories we
consider.  In Fig.~\ref{fig:convergence_spherical_sym}, we demonstrate
convergence of the spherically symmetric code.

\begin{widetext}
\begin{minipage}{\linewidth}
\begin{figure}[H]
\begin{center}
    \includegraphics[width=0.45\columnwidth,draft=false]{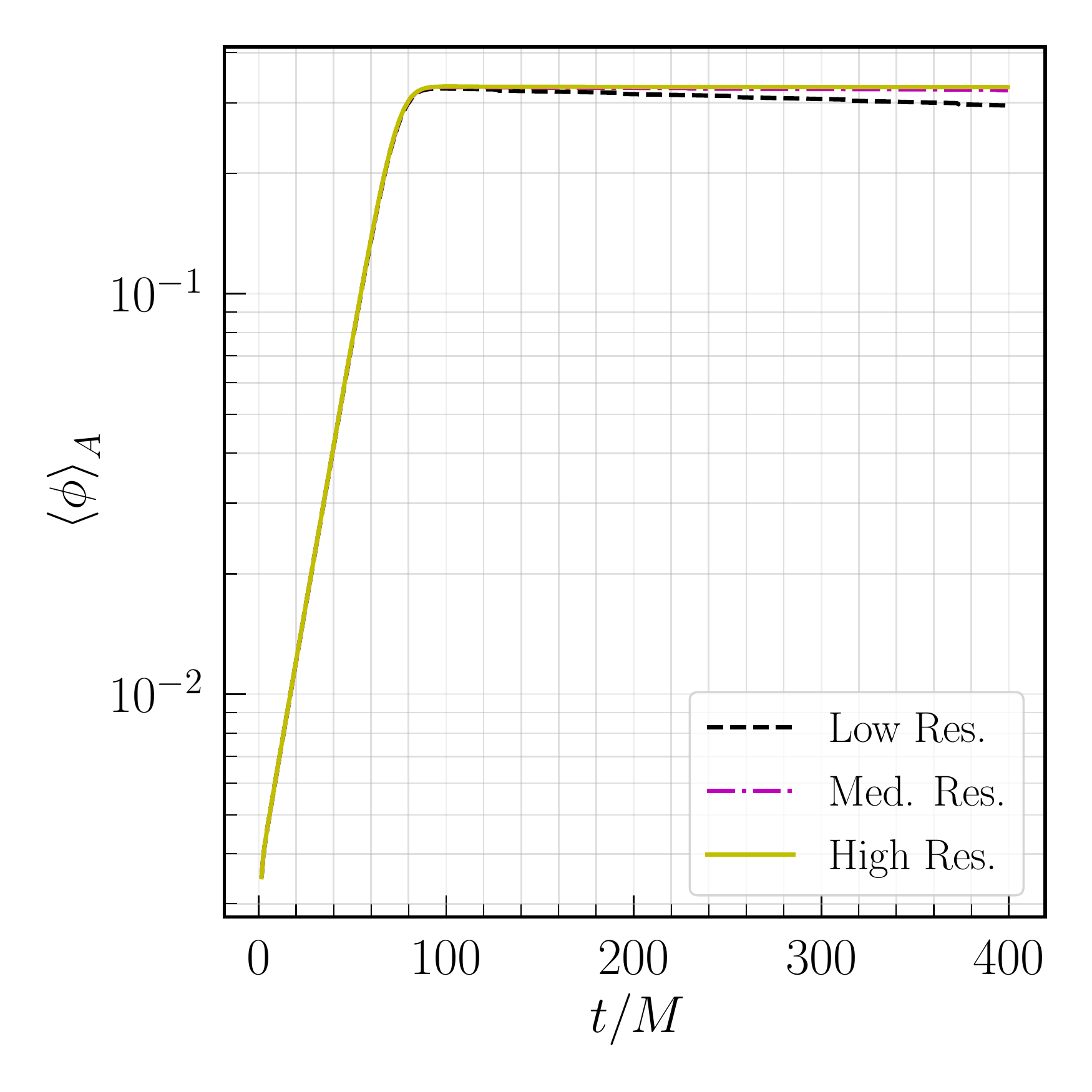}
    \includegraphics[width=0.45\columnwidth,draft=false]{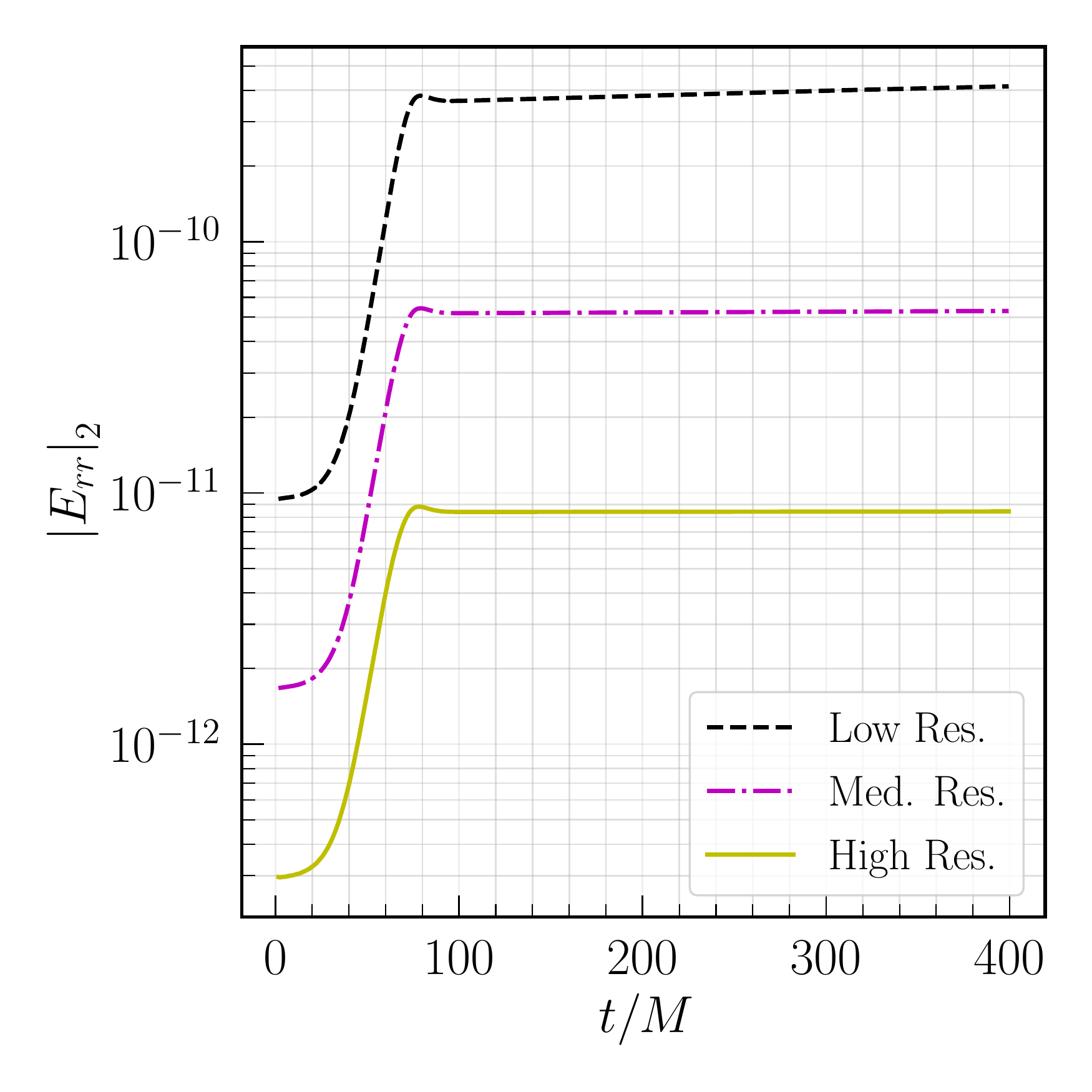}
\end{center}
\caption{
   We plot the value of the scalar field at the BH horizon,
   and the norm of the $rr$ component of the Einstein equations
   in spherical symmetry, for the case of exponential
   coupling Eq.~\eqref{eqn:exponential_coupling}, with $\lambda/M^2=0.8$.
   The slow decrease in the value of $\phi$ over time is due to the slow
   increase (due to truncation error) of the BH mass.
   The low, medium, and high resolutions are with radial grid
   resolution $nx=2^{12}+1$, $nx=2^{13}+1$, and $nx=2^{14}+1$,
   respectively.
   We see that we can both resolve the scalar
   field dynamics, and that the residuals are converging between
   second and fourth order, which is consistent with the second order
   method we use to solve for the metric variable and fourth
   order stencils we use to evolve the scalar field equations of motion
   (for more details on the numerical implementation,
   see Ref.~\cite{Ripley:2020vpk}).
\label{fig:convergence_spherical_sym}
}
\end{figure}
\end{minipage}
\end{widetext}


\end{document}